\documentclass[12pt]{article}
\renewcommand{\abstractname}{\ }
\begin{document}
\renewcommand{\abstractname}{\ }

\title{Modelling Quantum Mechanical Processes by Processes   Energy Distribution of Inner Oscillations of a Nanoparticle in the Phase Space}
\author{E. M. Beniaminov}
\date{}
\maketitle
\begin{abstract}
{We consider the problem of computing energy distribution of inner harmonic oscillations of a nanoparticle
in its phase space, when the particle moves in a medium in heat equilibrium under certain temperature.
It is assumed that the particle obeys the Brownian motion under the action of the medium and the force field
given by a potential function. In the present paper we provide and study an equation describing the problem,
generalizing the Klein--Kramers equation.
 It is shown that for large value of medium resistance, the process of energy distribution of inner harmonic oscillations
of the nanoparticle is represented as the composition of a rapid transition process and a slow process.
After the rapid transition process, the system goes to a quasi-stationary state.
The slow process is approximately described by the standard Schr\"odinger equation used for description of
quantum processes.
Thus, the process being studied can serve as a model of quantum processes.
}
\end{abstract}

{\bf Keywords:} quantum mechanics; Brownian motion; phase space; Klein--Kramers equation;
waves in phase space; asymptotic solutions.

\section{Introduction}
The purpose of the paper is to give an example of a mathematical model, based on natural assumptions, in which
the quantum or classical behavior of the modeled process occurs depending on the values of the parameters of the model.

In this paper we consider the mathematical model of distribution of certain characteristic (energy of inner
harmonic oscillations) of a nanoparticle in its phase space. It is assumed that the particle moves in a
medium and obeys the Brownian motion under the action of the medium and a field of external forces given by
certain potential function. We will be interested not in the probability distribution
of the position of the particle in the phase space (as in the standard problem on the Brownian motion) but in the
distribution of the energy of inner harmonic oscillations of the nanoparticle in its phase space.
It is also assumed that the frequency of inner oscillations of the nanoparticle is large.

Let us be more exact. In the paper, we consider a classical particle whose state is given by coordinates, momenta,
and certain parameters providing inner state of the particle.

The particle moves under the action of a force field in a medium being in heat equilibrium with certain temperature.

The medium resists to the movement of the particle with the force proportional to the velocity of the particle,
and changes randomly the velocity of the particle under collisions of the particle with the particles of the medium.
That is, the particle is in the Brownian motion.

It is assumed that the inner state of the particle is described by certain parameter whose value
oscillates harmonically with a large frequency $\omega>\!\!>1$ which is constant and does not change in the proper
time of the particle.

{\bf Problem:}
{\it Find the distribution of energy of inner harmonic oscillations of the particle in its phase space, for
the particle obeying the Brownian motion in the heat medium and action of external force field given
by a potential function. }

A state of such process is naturally described by the distribution of amplitudes and phases of inner oscillations
in the phase space, i.~e. a complex valued function on the phase space.

In the paper, we provide a mathematical model of this process in the form of the modified Klein--Kramers equation.
The Klein--Kramers equation \cite{kramers, van_kampen} describes the Brownian motion of the particle in the phase space.
Further, we study the constructed mathematical model depending on various values of the parameters of the model.
It is shown that in the case of large specific resistance of the medium, the process passes several stages in time.
During the first rapid stage the function giving the state of the process transfers to one of quasi-stationary states.
In the second slow stage the function evolves already in the subspace of quasistationary states
subject to the standard Schr\"odinger equation.

Further dissipation of the process leads to the fact that any superposition of eigenstates of the Hamilton operator
goes to one of eigenstates (the decoherence process). At the last stage, the mixed state of heat equilibrium
(the Gibbs state) arises, due to the heat action of the medium and transitions between eigenstates of the Hamilton
operator due to large random deviations.

If, on the contrary, the medium resistance per unit of mass of the particle is small, then it is shown that
in the considered model the density of energy distribution satisfies the classical Liouville equation, i.~e.
the process behaves as a classical system.

{\bf The main result:} {\it The presented process can model quantum processes depending on the values of the
parameters.}

\section{The mathematical model of the process}
Thus, in this paper we consider a particle moving in a heat medium. The states of the particle are determined by
classical coordinates $x\in R^3$ and momenta $p\in R^3$, and also by parameters determining the inner
state of the particle. It is assumed that the inner state performs small oscillations with a large frequency $\omega$,
and in the approximation being considered, it is described by a vector $\bar a\in V$ from a vector space $V$,
evolving according to the harmonic oscillations law with the frequency $\omega$.

It is assumed that under the motion of the particle the oscillating vector of inner state $\bar a\in V$
is transferred parallel to itself with a trivial connection. Hence, if we fix $\bar e$,
the direction of the vector $\bar a$, then the remaining degrees of freedom of the vector
$\bar a=\bar e A\cos(\omega \tau+ \alpha)$ oscillating in the proper time $\tau$, are given by the
amplitude $A$ and the phase $\omega \tau+ \alpha$.  The amplitude and the phase provide, in the standard way,
the complex number $\varphi=A\exp (- i(\omega \tau+ \alpha))\in {C}$, where $i$ is the imaginary unit.
The minus sign before the exponent is chosen so as to make the final expressions taking the form familiar in physics.
The complex number $\varphi\in C$ determines the inner state vector uniquely by the formula $\bar a=\bar e*Re(\varphi)$,
where $Re(\varphi)$ is the real part of the complex number $\varphi$.

Thus, under the assumptions considered, a state of the process at the moment of time $t$ is described by the
distribution $\varphi(x,p,t)\in C$, where $(x,p)\in R^6$ are the coordinates and the momenta of the particle,
and $\varphi$ is a complex number providing the inner state vector of the particle at a point of the phase space
of the particle at the moment of time $t$.

The quantity $|\varphi (x,p,t)|^2=A^2(x,p,t)$ is proportional to the energy distribution of inner oscillations of the
particle in the phase space at the moment $t$, where $|\varphi|$ denotes the absolute value of the complex number
$\varphi$, and $A$ is the amplitude of inner harmonic oscillations.
In the paper, we study the evolution process of the distribution $\varphi$ and respectively of $|\varphi|^2$.

Let us now proceed to the assumptions on actions on the particle forcing it to change its state.

The particle is acted on by an external force field given by the potential function $V(x)$.
Besides that, the particle moves in a medium being in heat equilibrium with the temperature $T$
and the medium resistance $\beta$. The particle performs the Brownian motion. If we denote by $f(x, p, t)$
the density of the probability distribution of the particle in the phase space at the moment of time $t$,
then this density $f(x,p,t)$ for our Brownian motion should satisfy the standard Klein--Kramers equation
\cite{kramers, van_kampen}:
\begin{equation}\label{eq_kramers}
\frac{\partial f}{\partial{t}}=\sum_{k=1}^{3}
\biggl(
\frac{\partial H}{\partial x_k} \frac{\partial f}{\partial p_k}-
 \frac {\partial H}{\partial p_k} \frac{\partial f}{\partial x_k}
\biggr)
+\frac{\beta}{m} \sum_{k=1}^{3}\frac{\partial }{\partial{p_k}}\biggl (p_k f
+k_B Tm\frac{\partial{f} }{\partial{p_k}}\biggr) ,
\end{equation}
where $H= c\sqrt {m^2c^2 +p^2} +V(x)$ is the Hamilton function;
$m$ is the mass of the particle;
$k_B$ is the Boltzmann constant.

Thus, the origin of the vector of inner state of the particle performs the Brownian motion
given by equation~(\ref{eq_kramers}).

Let us now consider how the phase of the inner state vector of the particle evolves.

It is assumed that the phase of the inner state vector of the particle moves with a constant large velocity
$\omega$ in the coordinate system related with the moving particle. That is, in this coordinate system
the complex number $\varphi$, corresponding to the inner state vector of the particle, evolves
according to the formula $\varphi=\varphi_0\exp(-i\omega\tau)$, where $i$ is the imaginary unit, $\tau$
is the proper time of the particle.

It is also assumed that the velocity $\omega$ is so large that the effects of special relativity
can influence on the phase of the inner state vector of the particle even for small velocities
of the motion of the particle itself.

Let us express the proper time $\tau$ of the particle, i.~e. the time in the coordinate system
related to the moving particle, through the time $t$ in the stationary coordinate system
by formulas of special relativity.
If a particle with the coordinates $x=(x_1,  x_2,  x_3)$  moves with
the velocity $v=(v_1,   v_2,   v_3)$,   then, according to the formulas of special relativity,
the proper time is expressed through the time $t$ of the observer by the following formula:
\begin{eqnarray}\label{tauv}
\tau=\frac{t-  (x v /{c^2})}{\sqrt{1-({v^2}/{c^2}})} ,\ \mbox{respectively,}\ \ \
d\tau=\frac{dt-   (v /{c^2})dx}{\sqrt{1-({v^2}/{c^2}})},
\end{eqnarray}
where $\   x v =x_1v_1+x_2v_2+x_3v_3$  is the scalar product of the vectors $x$ and $v$; $c$
is the light velocity. Note that according to special relativity theory, the proper time of the particle
is not changed under the change of its velocity. Hence the latter formula has no summand with
the factor $dv$.

     For the free particle with the momentum $p=(p_1, p_2, p_3)$ and the rest mass
$m$, the energy $ E=c\sqrt{p^2+m^2c^2}$ and, respectively,
\begin{equation}\label{v}
v =\frac{\partial E}{\partial p}= \frac{pc}{\sqrt{p^2+m^2c^2}},\ \ \ \ \
\sqrt{1-\frac{v^2}{c^2}}  =  \frac{mc}{\sqrt{p^2+m^2c^2}}=
\frac{mc^2}{E}.
\end{equation}

   Substituting these expressions into (\ref{tauv}),  after transformations we obtain:
\begin{equation}\label{tau}
 \tau=\frac{Et- x p}{mc^2} \ \ \ \mbox{и}\ \ \ d\tau=\frac{Edt- pdx}{mc^2}.
\end{equation}

The next important assumption is that formula~(\ref{tau}) holds not only for the free motion of the particle
but also for motion in the potential force field (at least in the case when the potential energy is
much smaller than the rest energy of the particle, i.~e. when $V(x)<\!\!< mc^2$), if instead of $E$
we substitute into this formula the Hamilton function $H=E+V(x)$. We obtain:
\begin{equation}\label{tauH}
 \tau=\frac{Ht- x p}{mc^2}  \ \ \ \mbox{and}\ \ \ d\tau=\frac{Hdt- pdx}{mc^2}.
\end{equation}

Let us consider the distribution in the phase space of complex vectors $\varphi(x,p,t)$,
whose arguments change with constant velocity $\omega$ in the proper time. If $\varphi(x,p,t)=\varphi_0$, for $\tau=0$,
then
\begin{equation}
 \varphi(x,p,t)=\varphi_0 \exp(-i\omega\tau).
\end{equation}
Substituting formula (\ref{tau}) into this expression and denoting
\begin{equation}\label{hbar}
\hbar\stackrel{def}{=} mc^2/\omega,
\end{equation}
we obtain
\begin{equation}\label{varphi_Htxp}
 \varphi(x,p,t)\!=\!\varphi_0 \exp(-i\omega\tau)\!=\!
\varphi_0 \exp\biggl(\!\frac{-i\omega(Ht- x p)}{mc^2}\!\biggr)\!=\!
\varphi_0 \exp\biggl(\!\frac{-i(Ht- x p)}{\hbar}\!\biggr).
\end{equation}

Denote by $ D_{\displaystyle x_k}$ and $ D_{\displaystyle p_k}$,  for $k=1,2,3$, the differentiation
operators of $\varphi(x,p,t)=\varphi_0 \exp(-i\omega\tau)$ under infinitely small parallel transport (shift) of
$\varphi$
by $dx_k$ and $dp_k$, respectively, without the change of the proper time $\tau$.
Formulas (\ref{tau}) and (\ref{varphi_Htxp}) imply that
\begin{equation}\label{D}
 D_{\displaystyle x_k}= {\partial}/{\partial{x_k}}-{ip_k}/{\hbar}\ \ \mbox{and}\ \
D_{\displaystyle p_k}= {\partial}/{\partial{p_k}},\ \ \mbox{where}\ \  k=1, 2, 3.
\end{equation}

Note also that such coordinate and momenta shift operators do not commute.
The commutators of their differential operators read as follows:
    $$ \left[D_{\displaystyle p_k},D_{\displaystyle x_k}\right]=
       -{i}/{\hbar}\mbox{  and  }
     \left[D_{\displaystyle p_k},D_{\displaystyle x_j}\right]=0,
       \mbox{ where } k\neq j \mbox{ and }  k,j=1,2,3.$$
Thus, such coordinate and momenta shifts of the wave functions $\varphi$ on the phase space
realize a representation of the Heisenberg group. This representation was considered by E. Prugovecki in the paper
\cite{prugovecki}.

Now we are completely ready to present the modified Klein--Kramers equation modelling the studied
process of distribution of amplitudes and phases of inner harmonic oscillations in the phase space.
We have seen that the state of the process at each moment of time $t$ is given by a complex valued function
$\varphi(x, p, t)$ on the phase space $(x,p) \in R^{6}$.
The energy distribution of these harmonic oscillations in the phase space is proportional to $|\varphi(x, p, t)|^2$,
and evolution of the function $\varphi(x, p, t)$ in time is given by the following modified Klein--Kramers equation:
\begin{eqnarray}\label{eq_diff0}
\frac{\partial \varphi}{\partial{t}}&=&\sum_{k=1}^{3}
\biggl(
\frac{\partial H}{\partial x_k}D_{\displaystyle x_k}\varphi-
 \frac {\partial H}{\partial p_k}D_{\displaystyle p_k} \varphi
\biggr)
-\frac{i}{\hbar} H\varphi+\nonumber \\
&&+\frac{\beta}{m} \sum_{k=1}^{3}D_{\displaystyle p_k}\biggl (i\hbar D_{\displaystyle x_k}{\varphi}
+k_B TmD_{\displaystyle p_k}\varphi\biggr),
\end{eqnarray}
where $\hbar =mc^2/\omega$.

The modified Klein--Kramers equation is obtained from the Klein--Kra\-mers equation (\ref{eq_kramers})
by replacement of the operators ${\partial}/{\partial x_k}$ and
 ${\partial}/{\partial p_k}$ by the differentiation operators (\ref{D})
 $D_{\displaystyle x_k}$ and $D_{\displaystyle p_k}$, respectively, by addition to the right hand side
 of the summand $-({i}/{\hbar})H\varphi$ and the replacement in the diffusion operator of multiplication
 operator of the function $\varphi $ by $p_k$
by the action of the operator $i\hbar D_{\displaystyle x_k}=(p_k+i\hbar \partial/{\partial x_k})$ on the function
$\varphi $.

The adding of the summand $-({i}/{\hbar})H\varphi$ is related to the fact that $\varphi$
describes the harmonic oscillations of the particle at the point $(x, p)$ with the frequency $\omega$ in the form
$\varphi=\varphi_0 \exp(-i\omega \tau)$ in the proper time $\tau$ of the particle or in the form
$d\varphi=-i\omega \varphi d\tau$. And according to formula (\ref{tauH}) one has $d\tau=(H/(mc^2))dt$ for $dx=0$.

After substitution into (\ref{eq_diff0}) of expressions (\ref{D}) for $D_{\displaystyle x_k}$ and
$D_{\displaystyle p_k}$,
the modified Klein--Kramers equation is represented in the following form:
\begin{equation}\label{eq_diff}
\frac{\partial\varphi}{\partial{t}}=A\varphi
+\gamma B{\varphi},
\end{equation}
\begin{equation}\label{def_A}
\mbox{где }\ \ \ \ \ \ A\varphi =\sum_{k=1}^{3}
\biggl(
\frac{\partial H}{\partial x_k} \frac{\partial\varphi}{\partial p_k}-
\frac {\partial H}{\partial p_k}
 \biggl(\frac{\partial }{\partial x_k}-\frac{i p_k}{\hbar}\biggr)\varphi
\biggr)
-\frac{i}{\hbar} H\varphi \ \ \ \ \ \ \
\end{equation}
\begin{equation}\label{def_B}
\mbox{и }\ \ \ \ \ B{\varphi}=
\sum_{k=1}^{3}\frac{\partial}{\partial{p_k}}\left( \biggl( p_k +i\hbar \frac{\partial}{\partial{x_k}}\biggr){\varphi}
+k_BTm\frac{\partial{\varphi} }{\partial{p_k}}\right); \ \ \ \ \ \ \ \nonumber
\end{equation}
$H(x, p)=c\sqrt{m^2c^2+p^2}\approx mc^2+p^2/(2m)$ is the Hamilton function of the system;
$i$ is the imaginary unit;
$\hbar$ by definition equals $mc^2/\omega$;
the parameter $\gamma=\beta /m$ is the specific resistance coefficient of the medium, i.~e. the
medium resistance coefficient
$\beta$, per unit of mass $m$ of the particle; $k_B$ is the Boltzmann constant; $T$ is the temperature of the medium
in which the particle moves.

The modified Klein--Kramers equation has been considered in \cite{ben2011, ben2014, ben2015}.
For the sake of completeness we shall present some results of these papers in the present paper.

Let us first consider the case when $\gamma=\beta/m $ is a large quantity, i.~e. when the
contribution of the operator $B$ into the general process of evolution of the wave function is large.
The main result obtained under this assumption is that the motion described by equation~(\ref{eq_diff})
asymptotically decomposes into a rapid motion and a slow one. The result of the rapid motion is that the
arbitrary wave function $\varphi(x, p, 0)$
reaches at the time of order $1/\gamma $ the subspace of eigenfunctions of the operator $B$
with eigenvalue 0. This subspace is parameterized by functions $\psi(x)$, depending only on
coordinates $x$. The slow motion takes place already in the subspace of such functions.
That is, the rapid motion yields stationary solutions of the diffusion equation
\begin{equation}\label{eq_diffB}
\frac{\partial{\varphi}}{\partial{t}}=\gamma B\varphi=\gamma \sum_{k=1}^{3}\frac{\partial}{\partial{p_k}}\left( \biggl(
p_k +i\hbar \frac{\partial}{\partial{x_k}}\biggr){\varphi}
+k_BTm\frac{\partial{\varphi} }{\partial{p_k}}\right)=0.
\end{equation}
Let us state a more exact statement.

{\bf Theorem 1.} {\it Let $\varphi(x,p,0)$ be a complex valued function on the phase space whose
Fourier transform with respect to $x$ tends to 0 as $p \rightarrow \infty$. The solution $\varphi(x,p,t)$
of the diffusion equation~(\ref{eq_diffB}) exponentially in time with the index $-\gamma t$,
i.~e. at the time of order $1/\gamma $,
goes to the function $\varphi_0=P_0\varphi$, of the following form:
 \begin{eqnarray}\label{p_0}
\varphi_0=P_0\varphi &=&
\frac{1}{(2\pi{\hbar})^{3}}\biggl(\frac{k_BTm}{\pi \hbar^2 } \biggr)^{3/2}
\!\!\!\int\limits_{R^{3}}\!\psi (y) e^{-\frac{k_BTm(x-y)^2}{2\hbar^2}}e^{\frac{i  p(x-y)}{\hbar}} dy,\\
\label{psi}
&&\mbox{where}\ \ \psi (y) =\biggl(\frac{\pi \hbar^2 }{k_BTm} \biggr)^{3/2}\int\limits_{R^3}\!\!\ \varphi(y,p,0) dp.
\end{eqnarray}
The functions of the form~(\ref{p_0}) form a linear subspace of stationary functions for equation~(\ref{eq_diffB})
in the space of functions $\varphi(x, p)$. This subspace is parameterized by functions $\psi(y)$,
depending only on coordinates $y\in R^3$. The operator $P_0$ is the projection operator onto this subspace.
}

Proof of Theorem 1 is given in  Appendix 1 to this paper.

The constant before integral in formula~(\ref{psi}) is chosen so that the following equality holds:

$$\int_{R^6}|\varphi_0(x,p)|^2 dx dp=\int_{R^3} |\psi(y)|^2 dy.$$

The representation of the Galileo group on the subspace of functions of the form (\ref{p_0}), but
without using the medium temperature parameter in the formula, has been considered by E. Prugovecki
in \cite{prugovecki}, and the generalization of this representation has been used by him in \cite{prugovecki_book}
for unification of quantum mechanics and relativity theory.

{\it {\bf Theorem 2}.
The motion described by equation~(\ref{eq_diff}) decomposes asymptotically for large $\gamma$
into a rapid motion and a slow one.  After the rapid motion the function $\varphi(x, p, 0)$
goes at the time of order $1/\gamma $ to the function $P_0\varphi$ from Theorem~1.

The slow motion starting with the function $P_0\varphi$ of the form~(\ref{p_0}) with nonzero
function $\psi(y)$ goes in the subspace of such functions, and is parameterized by the function
$\psi(y,t)$ depending on time. This function $\psi(y,t)$ satisfies the Schr\"odinger equation of the form
$ i\hbar {\partial \psi}/{\partial t} = \hat{H}\psi$, where
\begin{equation}\label{hatH}
{\hat H}\psi =-\sum_{k=1}^{3} \frac{\hbar^2}{2m}\frac{\partial^2 \psi }{\partial{y^2_k}}
+V(y) \psi+mc^2\psi-\frac{3kT}{2}\psi + O(1/\gamma);
\end{equation}
${\hat H}$ is an operator differing from the standard Hamilton operator by constant summands.
}

Proof of Theorem~2 is given in Appendix~2 to this paper.

Note that in this model, reversibility of the quantum process given by the operator (\ref{hatH})
is the result of description of a non-invertible process given by equations~(\ref{eq_diff}, \ref{def_A}, \ref{def_B}),
but in the 0-approximation with respect to the parameter $1/\gamma$.

We shall not use what follows, but one would like to mention that in \cite{ben2015} one considers
the approximation of the operator $ \hat{H}$ from Theorem~2 with precision up to $O(1/\gamma^2)$, and one states
that in the motion described by equation~(\ref{eq_diff}), there is a more slow motion determined by dissipation of the
process, after which any superposition of eigenstates of the Hamilton operator goes to one of
eigenstates.  This motion corresponds in quantum mechanics to the decoherence process \cite{zurek}.
In the same paper an assumption has been made that the mixed heat equilibrium state (the Gibbs state)
arises in the additional much more slow motion of this process due to the heat action of the medium
and the transitions between
the eigenstates of the Hamilton operator due to large random deviations.

There is a lot of papers devoted to the study of processes described by the Klein--Kramers equation.
The remarkable surveys of this subject are the papers \cite{melnikov, berglund}. It is typical for such processes
that they are represented in the form of rapid and slow motions, and also that jumps between
quasi-stationary states occur.

Let us now consider the case when $\gamma=0$. In this case, the modified Klein--Kramers equation
 (\ref{eq_diff}, \ref{def_A}, \ref{def_B}) is represented as follows:
\begin{equation}\label{eq_diffA}
 \frac{\partial\varphi}{\partial t} =\sum_{k=1}^{3}
\biggl(
\frac{\partial H}{\partial x_k} \frac{\partial\varphi}{\partial p_k}-
\frac {\partial H}{\partial p_k}
 \biggl(\frac{\partial }{\partial x_k}-\frac{i p_k}{\hbar}\biggr)\varphi
\biggr)
-\frac{i}{\hbar} H\varphi.
\end{equation}

Let us introduce the notation $\rho=|\varphi|^2=\varphi\varphi^*$, where the sign $*$ denotes the operation
of complex conjugation. According to our definitions, the function $\rho$ describes the distribution
of energy of inner harmonic oscillations of the particle in the phase space.

{\bf Theorem 3.} {\it If the function $\varphi$ satisfies the modified Klein--Kramers equation for $\gamma=0$,
i.~e. equation~(\ref{eq_diffA}), then $\rho=\varphi\varphi^*$ satisfies the classical Liouville equation:
\begin{equation}\label{eq_rho}
 \frac{\partial\rho}{\partial t} =\sum_{k=1}^{3}
\biggl(
\frac{\partial H}{\partial x_k} \frac{\partial\rho}{\partial p_k}-
\frac {\partial H}{\partial p_k}
 \frac{\partial \rho}{\partial x_k}
\biggr).
\end{equation}
}
{\bf Proof.} Let us apply to both sides of equation (\ref{eq_diffA}) the operation of complex conjugation. We obtain:
\begin{equation}\label{eq_diffA'}
 \frac{\partial\varphi^*}{\partial t} =\sum_{k=1}^{3}
\biggl(
\frac{\partial H}{\partial x_k} \frac{\partial\varphi^*}{\partial p_k}-
\frac {\partial H}{\partial p_k}
 \biggl(\frac{\partial }{\partial x_k}+\frac{i p_k}{\hbar}\biggr)\varphi^*
\biggr)
+\frac{i}{\hbar} H\varphi^*.
\end{equation}

By the property of derivative of a product, we have:
$$\frac{\partial\rho}{\partial t}=\frac{\partial(\varphi\varphi^*)}{\partial t}=
\varphi^*\frac{\partial\varphi}{\partial t}+\varphi\frac{\partial\varphi^*}{\partial t}.
$$
If we subatitute into this expression instead of ${\partial\varphi}/{\partial t}$ and
${\partial\varphi^*}/{\partial t}$ their expressions by formulas (\ref{eq_diffA})
and (\ref{eq_diffA'}) respectively, open the brackets, join the similar terms and join in groups,
taking ${\partial H}/{\partial x_k}$ and ${\partial H}/{\partial p_k}$
out of brackets, then we obtain the required expression (\ref{eq_rho}).

Theorem~3 implies that if the particle in the medium does not interact with the medium $(\gamma=0)$,
then the energy distribution of inner harmonic oscillations of the particle moves along the
classical trajectories of the particle. That is, in this model for an isolated particle
not interacting with the medium, quantum effects do not arise.

\section{Conclusion}
In the paper we considered a mathematical model of the process of energy distribution in the phase space
for inner harmonic oscillations of a particle in the Brownian motion. A state of such process
is described by a complex valued function on the phase space of the particle.

It is shown that in the model considered, the process for $\gamma=\beta/m>\!\!>1$ decomposes
into a rapid (of order $1/\gamma$) and a slow motions. After the rapid motion the system,
starting with a state represented by arbitrary function on the phase space,
goes to states represented by functions belonging to certain distinguished subspace.
The elements of this distinguished subspace correspond to functions depending only on coordinates
(or only on momenta).
The slow motion, starting with a nonzero function from this subspace, goes inside the subspace and is described by the
usual Schr\"odinger equation.

If, on the contrary, the resistance of the medium per unit of mass of the particle is small, then in the considered
model, the energy distribution density of oscillations satisfies the classical Liouville equation,
which corresponds to the classical process.

Among the conclusions of this paper, one should mention the following ones.

In the presented model the processes described by the usual equations of quantum mechanics,
arise as approximate descriptions, resulting from asymptotics of the processes studied here,
of energy distribution of inner harmonic oscillations of the particle in its phase space
under interaction with medium. Therefore, such processes can serve as a convenient model
for processes of quantum mechanics.

The invertibility of quantum processes in this model is a result of approximate description of the process
in 0-approximation with respect to the parameter $1/\gamma$.

And, finally, it is shown that if in this model one excludes interaction of the particle with the medium,
then the process is described by the laws of classical mechanics.

Note also that in the presented model, one can consider the process with a small value of the parameter $\gamma$,
when this process is already not exactly described by the laws of classical mechanics, but is not yet described by
the laws of quantum mechanics.

It is also important to note that in the proposed model the value of the parameter $\hbar\stackrel{def}{=}mc^2/\omega$
is determined by the mass of the particle and the frequency of its inner oscillations and, in the general case,
can be arbitrary.
Moreover, for interacting particles the value $\hbar$ for each particle can be different.
That is, the universality of the Planck constant $\hbar$ in quantum mechanics is not derived in this
model, and for such derivation one needs additional conditions.

The approach proposed in this paper reminds the approach proposed in the paper \cite{comisar},
where one considered the model of quantum mechanics in the form of Brownian motion,
but it uses the diffusion coefficient in the form of a complex number,
and arising of complex numbers in the model has no substantiation.

\section*{Appendix}

\subsection*{Appendix 1. Proof of Theorem 1}
Let us substitute into equation~(\ref{eq_diffB})  the presentation of $\varphi(x,p,t)$ as a Fourier integral of the
following form:
\begin{equation} \label{fur}
  \varphi(x,p,t)={\cal F}_\hbar \tilde\varphi \stackrel{def}{=}\frac{1}{(2\pi \hbar)^{3/2}}
\int_{R^3}\tilde\varphi(s,p,t)e^{i s x/\hbar}ds,
\end{equation}
\begin{equation} \label{fur_1}
\mbox{where  }\ \
  \tilde\varphi(s,p,t)={\cal F}_\hbar^{-1}\varphi \stackrel{def}{=}\frac{1}{(2\pi \hbar)^{3/2}}
\int_{R^3}\varphi(x,p,t)e^{-{i s x/\hbar }}dx.
\end{equation}

We obtain that $ \tilde\varphi(s,p,t)$ satisfies the following equation:
\begin{equation}\label{eq_L_tilde}
\frac{\partial\tilde\varphi}{\partial{t}}=
\gamma \sum_{k=1}^{3}\frac{\partial}{\partial{p_j}} \biggl(( p_j -s_j){\tilde\varphi}
+k_BTm\frac{\partial{\tilde\varphi} }{\partial{p_j}}\biggr).
\end{equation}

The operator in the right hand side of this equation is well known (see, for example, \cite{kamke}).
This operator has a complete set of eigenfunctions in the space of functions tending to zero as $|p|$
tends to infinity.
The eigenvalues of this operator are the non-positive integers multiplied by $-\gamma$, i.~e.
$0, -\gamma, -2\gamma,...$
The eigenvalue 0 corresponds to eigenfunctions of the form
$$\tilde\varphi_0(s, p)= \frac{1}{(2\pi)^{3/2}}\tilde\psi(s)\exp\biggl({-\frac{(p-s)^2}{2k_BTm}}\biggr),$$
where $\tilde\psi(s)$ is an arbitrary complex valued function of $s\in R^3$.

The remaining eigenfunctions are obtained as derivatives of the functions $\tilde\varphi_0(s, p)$ with respect to $p$,
and have the eigenvalues $-\gamma, -2\gamma,  ...$ respectively, depending on the order of the derivative,
and the projector $P_0$ onto the subspace of eigenfunctions with eigenvalue 0, up to a constant $C$,
reads as follows:
\begin{equation}\label{P_0_tilde}
 \tilde\varphi_0(s, p)=P_0 \tilde\varphi=\frac{C}{(2\pi k_B Tm)^{3/2}}\tilde\psi(s) e^{-\frac{(p-s)^2}{2k_BTm}},\ \
\mbox{where} \ \ \tilde\psi(s)=\frac{1}{C}\int\limits_{R^3}\tilde\varphi (s,p)dp.
\end{equation}
Hence, considering equation~(\ref{eq_L_tilde}) in the basis of eigenfunctions,
we obtain that each solution $ \tilde\varphi(s,p,t)$ of this equation exponentially in time with index
$-1/\gamma$ tends to a stationary solution of the form $\tilde\varphi_0$.  Thus, using
the presentation~(\ref{fur})  of the function $\varphi_0(x,p,t)$ through $\tilde\varphi_0(s,p,t)$,
we obtain that stationary solutions $\varphi_0(x,p)$ of equation~(\ref{eq_diffB})  read as follows:
$$
\varphi_0(x,p)=\frac{1}{(2\pi\hbar)^{3/2}}\frac{C}{(2\pi k_B Tm)^{3/2}}
\int_{R^3}\tilde \psi(s) e^{-\frac{(p-s)^2}{2k_BTm}}
 e^\frac{i s x}{\hbar}ds.
$$

Let us represent the function $\tilde\psi(s)$, in its turn, in the form of the inverse Fourier transform:
$$
  \tilde\psi (s) =\frac{1}{(2\pi \hbar)^{3/2}}
\int_{R^3}\psi (y)e^{-\frac{i s y}{\hbar}}dy.
$$
Substituting this expression into the previous expression and integrating over $s$, we obtain:
$$
\varphi_0(x,p)=P_0\varphi(x,p)=\frac{1}{(2\pi\hbar)^{3}}\frac{C}{(2\pi k_B Tm)^{3/2}}
\int_{R^{6}}\psi (y) e^{-\frac{(p-s)^2}{2k_BTm}}
e^\frac{i s (x-y)}{\hbar}ds\ dy
$$
$$
=\frac{1}{(2\pi\hbar)^{3}}C
\int_{R^3}\psi (y) e^{-\frac{k_BTm(x-y)^2}{2\hbar^2}}
e^\frac{i p(x- y) }{\hbar} dy,
\ \ \mbox{where} \ \ \psi(s)=\frac{1}{C}\int\limits_{R^3}\varphi (s,p)dp,$$
which coincides with formula (\ref{p_0}) of Theorem 1 up to the value of the constant $C$.
The value of the constant $C$ can be arbitrary, but here it is chosen from the following equality:

$$\int_{R^6}\varphi_0 \varphi_0^* dx dp=\int_{R^3} \psi\psi^*dy. $$

Simple computations show that in this case one has
$$
C=\biggl(\frac{k_BTm}{\pi \hbar^2 } \biggr)^{3/2}.
$$
Note that the integral of the obtained expression $P_0\varphi$ with respect to $dp$ yields $C\psi(x)$.
This implies that $P_0^2=P_0$.  That is, $P_0$ is a projection operator.

Theorem 1 is proved.

\subsection*{Appendix 2. Proof of Theorem 2}
Consider the process given by equations~(\ref{eq_diff}, \ref{def_A}, \ref{def_B}). For large $\gamma$
the main contribution to the right hand side of this equation gives the operator $B$.
By Theorem 1, if one does not take into account the contribution of the operator $A$,
then a state of the process after the time of order $1/\gamma$ will be described by a function of
the form~(\ref{p_0}).

Let $\varphi_0(x,p,t)$ be the function of the form~(\ref{p_0}) corresponding to the function $\psi(y,t)$.
Let us substitute this expression $\varphi_0(x,p,t)$ into equations~(\ref{eq_diff}, \ref{def_A}, \ref{def_B})
instead of $\varphi$. Let us take into account that $B\varphi_0=0$, by Theorem~1.
Then let us apply to both sides of the obtained equation the operation given by equality~(\ref{psi}),
i.~e. let us take the integral of both parts over $p$, multiplying it by the constant
standing before the integral in formula~(\ref{psi}). We have:
$$
\frac{1}{(2\pi\hbar)^{3}}\!\!\int\limits_{R^{6}}\!\!\frac{\partial \psi (y,t)}{\partial t}
e^{-\frac {k_BTm(x-y)^2}{2\hbar^2}}
e^{i p(x- y)/\hbar } dydp=\ \ \ \ \ \ \ \ \ \
$$
$$\ \ \ \ \ \ \ \ \ \ \ \ \ \ \ \ \ \ \ \ \ =\frac{1}{(2\pi\hbar)^{3}}
\!\!\int\limits_{R^{6}}\!\!A\psi (y,t)  e^{-\frac {k_BTm(x-y)^2}{2\hbar^2}}
e^{i p(x- y)/\hbar } dydp.
$$
Let us integrate the left hand side of this equality over $p$ and over $y$; noting that one has the delta function there,
we obtain:
$$
\frac{\partial \psi (x,t)}{\partial t}=\frac{1}{(2\pi\hbar)^{3}}
\!\!\int\limits_{R^{6}}\!\!A\psi (y,t)  e^{-\frac {k_BTm(x-y)^2}{2\hbar^2}}
e^{i p(x- y)/\hbar } dydp.
$$
Taking into account expression~(\ref{def_A}) for the operator $A$, we obtain from the latter equality and
from additivity of integral:
\begin{eqnarray}\label{Hfull}
\frac{\partial \psi }{\partial t}\!\!\!\!&=&\!\!\!\!\frac{1}{(2\pi\hbar)^{3}}
\!\!\int\limits_{R^{6}}\!\!\left(\sum_{j=1}^{3}
\biggl(
\frac{\partial V}{\partial x_j} \frac{\partial}{\partial p_j}-
  \frac{p_j}{m} \frac{\partial }{\partial x_j}
\biggr)
-\frac{i}{\hbar}
\biggl(mc^2+V-\sum_{j=1}^{3}\frac{p_j^2}{2m} \biggr)\!\!\right)\nonumber\\
\!\!\!&\times&\!\!\!\psi (y,t)  e^{-\frac {k_BTm(x-y)^2}{2\hbar^2}}
e^{{i p(x- y)}/{\hbar }} dydp=
I_1+I_2+I_3+I_4,\ \mbox{where}
\end{eqnarray}
\begin{eqnarray}\label{def_I_1}
I_1\!\!&=&\!\!\frac{1}{(2\pi\hbar)^{3}}\int\limits_{R^{6}}\sum_{j=1}^{3}\frac{\partial V(x)}{\partial x_j}
\frac{\partial}{\partial p_j}
\left(\psi (y,t)  e^{-\frac {k_BTm(x-y)^2}{2\hbar^2}}
e^{{i p(x- y)}/{\hbar }} \right)dydp;\\
I_2\!\!&=&\!\!-\frac{1}{(2\pi\hbar)^{3}}\int\limits_{R^{6}}\sum_{j=1}^{3}\frac{p_j}{m}
\frac{\partial }{\partial x_j}\left(
\psi (y,t)  e^{-\frac {k_BTm(x-y)^2}{2\hbar^2}}
e^{{i p(x- y)}/{\hbar }} \right)dydp;\label{def_I_2}\\
I_3\!\!&=&\!\!-\frac{i}{\hbar}\frac{1}{(2\pi)^{3}}\int\limits_{R^{6}}\biggl(mc^2+V(x)\biggr)
\psi (y,t)  e^{-\frac {k_BTm(x-y)^2}{2\hbar^2}}
e^{{i p(x- y)}/{\hbar }} dydp;\label{def_I_3}\\
I_4\!\!&=&\!\!\frac{i}{\hbar}\frac{1}{(2\pi)^{3}}\int\limits_{R^{6}}\sum_{j=1}^{3}\frac{p_j^2}{2m}
\psi (y,t)  e^{-\frac {k_BTm(x-y)^2}{2\hbar^2}}
e^{{i p(x- y)}/{\hbar }} dydp.\label{def_I_4}
\end{eqnarray}

Consider the integral $I_1$ given by expression~(\ref{def_I_1}). Let us exchange summation and integration,
take out of the sign of integral the expressions not depending on integration variables, compute derivatives
with respect to $p_j$,  and integrate the remaining integrals over $p$ and $y$.
We obtain:
\begin{eqnarray}\label{res_I_1}
I_1=\frac{1}{(2\pi\hbar)^{3}}\sum_{j=1}^3\frac{\partial V(x)}{\partial x_j}\int\limits_{R^{6}}
\psi (y,t)  \frac{i(x_j-y_j)}{\hbar}
e^{-\frac {k_BTm(x-y)^2}{2\hbar^2}}
e^{{i p(x- y)}/{\hbar }} dydp=0
\end{eqnarray}

Consider the integral $I_2$ given by expression~(\ref{def_I_2}). Let us exchange summation and integration in it,
take the factor $1/m$ out of the sum and integral signs,
take the derivatives with respect to $x_j$ out of the sign of integral, replace the expressions
$p_j \exp(ip(x-y)/\hbar)$ by the equal expressions
$i\hbar\partial \ \exp(ip(x-y)/\hbar)/(\partial y_j)$, and integrate the obtained integrals by parts. We have:
\begin{eqnarray}\label{res_I_2}
I_2&=&-\frac{1}{m(2\pi\hbar)^{3}}\sum_{j=1}^{3}\frac{\partial }{\partial x_j}\int\limits_{R^{6}}
\psi (y,t) e^{-\frac {k_BTm(x-y)^2}{2\hbar^2}}i\hbar\frac{\partial  }{\partial y_j}\left (e^{{i p(x- y)}/{\hbar }}\right)
dydp\nonumber\\
&=&\frac{i\hbar}{m(2\pi\hbar)^{3}}\sum_{j=1}^{3}\frac{\partial }{\partial x_j}\int\limits_{R^{6}}
    \frac{\partial \psi (y,t)}{\partial y_j}e^{-\frac {k_BTm(x-y)^2}{2\hbar^2}}
e^{{i p(x- y)}/{\hbar }} dydp\nonumber\\
&+&\!\!\!\!\frac{i\hbar}{m(2\pi\hbar)^{3}}\sum_{j=1}^{3}\frac{\partial }{\partial x_j}\int\limits_{R^{6}}
    \psi (y,t)\frac {k_BTm}{\hbar^2}(x_j-y_j)e^{-\frac {k_BTm(x-y)^2}{2\hbar^2}}
e^{{i p(x- y)}/{\hbar }} dydp\nonumber\\
&=&\frac{i\hbar}{m}\sum_{j=1}^{3}\frac{\partial^2  \psi (x,t)}{\partial x_j^2}.
\end{eqnarray}

Consider the integral $I_3$ given by expression~(\ref{def_I_3}). Let us take out of the integral sign in it
the expressions not depending on integration variables, and integrate the remaining integral over $p$ and $y$.
We obtain:
\begin{eqnarray}\label{res_I_3}
I_3&=&-\frac{i}{\hbar}(mc^2+V(x)) \frac{1}{(2\pi\hbar)^{3}}\int\limits_{R^{6}}
\psi (y,t) e^{-\frac {k_BTm(x-y)^2}{2\hbar^2}}
e^{{i p(x- y)}/{\hbar }} dydp\nonumber\\
&=&-\frac{i}{\hbar}(mc^2+V(x))\psi (x,t).
\end{eqnarray}

Consider the integral $I_4$ given by expression~(\ref{def_I_4}). Let us exchange summation and integration in it,
take the factor $1/2m$ out of the sum and integral signs, replace the expression $p_j^2 \exp(ip(x-y)\hbar)$
by the equal expression through the second derivatives with respect to $y_j$ of the form
$\hbar^2 \partial^2 \exp(ip(x-y)\hbar)/ \partial y_j^2$,
and integrate the obtained integrals by parts with respect to $y_j$. We have:
\begin{eqnarray}\label{res_I_4}
I_4&=&-\frac{i}{2m\hbar}\frac{1}{(2\pi\hbar)^{3}}\sum_{j=1}^{3}\int\limits_{R^{6}}
\psi (y,t) e^{-\frac {k_BTm(x-y)^2}{2\hbar^2}}\frac{\hbar^2\partial^2  }{\partial y_j^2}e^{i p(x- y)/\hbar } dy dp
\nonumber\\
&=&-\frac{i\hbar}{2m}\frac{1}{(2\pi\hbar)^{3}}\sum_{j=1}^{3}\int\limits_{R^{6}}
    \biggl(\frac{\partial^2 \psi }{\partial y_j^2}+2\frac {k_BTm}{\hbar^2}(x_j-y_j)\frac{\partial \psi }{\partial y_j}
    \nonumber\\
&& +\biggl(\frac {k_BTm}{\hbar^2}\biggr)^2 (x_j-y_j)^2
\psi - \frac {k_BTm}{\hbar^2}\psi \biggr)
\times e^{-\frac {k_BTm(x-y)^2}{2\hbar^2}}
e^{{i p(x- y)}/{\hbar }} dydp\nonumber\\
&=&-\frac{i\hbar}{2m}\left(\sum_{j=1}^{3}\frac{\partial^2  \psi (x,t)}
{\partial x_j^2} -3\frac {k_BTm}{\hbar^2} \psi (x,t)\right).
\end{eqnarray}

Let us substitute the obtained expressions for the integrals $I_1$, $I_2$, $I_3$,  $I_4$
into equality~(\ref{Hfull}), join the similar terms, and multiply both parts of the equality by $i\hbar$.
We obtain,
$$
i\hbar\frac{\partial \psi }{\partial t}=
-\sum_{j=1}^{3}\frac{\hbar^2}{2m}\frac{\partial^2\psi}{\partial x_j^2}+
V\psi+ mc^2\psi-\frac{3k_BT}{2}\psi,
$$
which coincides with equality~(\ref{hatH}) required in Theorem~2.


\begin{thebibliography}{99}
\bibitem{kramers}
{H. A. Kramers,}
{\it Brownian motion in a field of force and the diffusion model of chemical reactions.}// Physica. 7, 284Ц304. (1940)
\bibitem{van_kampen}
{N.G. Van Kampen,} {Stochastic Processes in Physics and Chemistry.} North Holland, Amsterdam, 1981.
\bibitem{prugovecki}
{S.T. Ali, E. Prugovecki,} {\it Quantum statistical mechanics on stochastic phase space.}// Int. J. Theor. Phys. 16,
689--706 (1977).
\bibitem{prugovecki_book}
 {E. Prugovecki,} {Stochastic Quantum Mechanics and Quantum Spacetime Ц A Consistent Unification of Relativity and
 Quantum
 Theory Based on Stochastic Spaces.} D. Reidel Publishing Company, Boston, 1984.
\bibitem{ben2011}
{ E.M. Beniaminov,}
{\it Quantum Mechanics as Asymptotics of Solutions of
Generalized Kramers Equation.}// Electronic Journal of Theoretical Physics
(EJTP) 8, No. 25, 195-210 (2011).
\bibitem{ben2014}
{ E.M. Beniaminov,}
{\it Diffusion Scattering of Waves is a Model of Subquantum Level.}// Electronic Journal of Theoretical Physics (EJTP) 11,
No. 30, 35Ц48 (2014).
http://www.ejtp.com/articles/ejtpv11i30p35.pdf.
\bibitem{ben2015}
{ E.M. Beniaminov,}
{\it Scattering of Waves in the Phase Space, Quantum Mechanics, and Irreversibility.}// Electronic Journal of Theoretical
Physics (EJTP) 12, No. 32,  43Ц60 (2015).
http://www.ejtp.com/articles/ejtpv8i25p195.pdf.
\bibitem{zurek}
{W. H. Zurek,} {\it Decoherence and the transition from quantum to classical - REVISITED}
arXiv:quant-ph/0306072v1. 2003 (An updated version of PHYSICS TODAY, 44:36-44 (1991)).
\bibitem{melnikov}
 {V.I. Mel'nikov,} {\it The Kramers Problem: Fifty Years of Development}//
Physics reports (Review Section of Physics Letters) 299. Nos. 1\&2 (1991), 1 Ч71.
\bibitem{berglund}
{N. Berglund,} {\it Kramers' law: Validity, derivation and generalizations}// arxiv:1106.5799v2 (2013).
\bibitem {kamke}
{E.~Kamke,}  {\it
Differentialgleichungen:
Losungsmethoden und Losungen}, I, Gewohnliche Differentialgleichungen, B. G. Teubner, Leipzig, 1977.
\bibitem{comisar}{G.G. Comisar}{\it Brownian-Motion Model of Nonrelativistic Quantum Mechanics}// Physical Review 138,
N 5 B (1965), 1332--1337.
\end{thebibliography}
\end{document}